\documentclass[12pt, draftclsnofoot, onecolumn]{IEEEtran}
\usepackage{bbm}
\usepackage{amsfonts}
\usepackage{tipa}
\usepackage{amssymb}
\usepackage{mathrsfs}
\usepackage{stfloats}
\usepackage{color, soul}
\usepackage[lined,boxed,commentsnumbered]{algorithm2e}
\usepackage{graphicx}
\usepackage{subfigure}
\usepackage{amsfonts,amssymb,amsmath}
\usepackage{latexsym}
\usepackage{multirow}
\usepackage{hyperref}
\usepackage{epsfig,epstopdf}
\usepackage{array}
\newtheorem{thm}{Theorem}
\newtheorem{remark}{Remark}
\newtheorem{lemma}{Lemma}

\begin{document}
\title{QoS Guaranteed Energy Minimization Over Two-Way Relaying Networks}
\author{Zhi~Chen~\IEEEmembership{Member,~IEEE} and Pin-Han~Ho~\IEEEmembership{Senior Member,~IEEE,}
\thanks{Z. Chen and P. Ho are with the Department of Electrical and Computer Engineering, University of Waterloo, Waterloo, Ontario, Canada, N2L3G1. Emails: chenzhi2223@gmail.com; p4ho@uwaterloo.ca.
}}

\maketitle

\baselineskip 24pt
\begin{abstract}
\baselineskip 18pt
In this work, we consider an energy minimization problem with network coding over a typical three-node, two-way relaying network (TWRN) in the wireless fading environment, where two end nodes requires the same average exchange rates no lower than
a predefined quality-of-service (QoS) constraint.
To simplify the discussion, the selected network coding modes only include physical-layer network
coding (PNC) and the superposition coding based digital network coding (SPC-DNC). We first analyze their energy usages and then propose a optimal strategy, which can be implemented by switching between PNC and SPC-DNC for each channel realization.
An iterative algorithm is hence presented to find the optimal power allocations as well as optimal time splitting for both uplink and downlink transmissions. The conducted numerical study validates the performance improvement of the new developed strategy.

\end{abstract}

\IEEEpeerreviewmaketitle

\section{Introduction}
In recent decades, relaying transmission as well as network coding have attracted increasing attention as these two techniques can well exploit cooperative diversity/network coding gain
to improve network performance in terms of metrics of interest \cite{ahlswede2000network}-\cite{Fan'09_JSAC}. Two-way relay channel, a
typical application which jointly utilizes relays and
network coding, has been extensively studied in
\cite{oechtering2008stability}-\cite{narayanan2007joint}, where the throughput of DNC are studied in
\cite{Fan'14_two_way} and the
rate region of PNC are studied in
\cite{narayanan2007joint}.

Further, green communication has received increasing attention, as it introduces novel solutions to greatly reduce energy consumptions in communication systems designs. in the literature, numerous works studied reducing energy usage while still satisfying the QoS requirement for various types of communication networks, e.g., \cite{Phuyal'12} investigated an energy-aware transmission strategy in a multi-user multi-relay cellular network and \cite{Wang'06} discussed various energy-aware scheduling algorithms in wireless sensor networks.

In this work,
we are motivated to analytically analyze the energy usage
of PNC and the superposition-coding based DNC (SPC-DNC).
We then
find the decision criterion in selecting PNC or SPC-DNC in terms of minimizing energy usage for each channel realization. Further, a PNC/SPC-DNC switching strategy is designed to smartly select the energy-efficient strategy under fading channel
realizations, with the QoS requirement still being satisfied.
To better compare the two strategies, we focus on the end-to-end symmetric throughput scenario. However, our analysis can be naturally extended to asymmetric throughput case, and is omitted here due to the limited scope of this work.

\section{System Description}
In this work, a three-node, two-way relaying network over fading channels is studied. In this TWRN, the two source nodes, $S_1$ and $S_2$ want to exchange data through the aid of the relay node, $\tt R$. All nodes work in half-duplex mode and cannot transmit and receive simultaneously. The direct link between the two sources is assumed to be unavailable. The channel power gains of the $S_i$-$\tt R$ ($i=1,2$) link is denoted as $g_{ir}$ and that of the $\tt R$-$S_i$ is $g_{ri}$. The noise at each node is assumed to be additive white Gaussian noise with zero mean and unit variance.

\begin{figure}[!t]
\centering
\includegraphics[width=12cm]{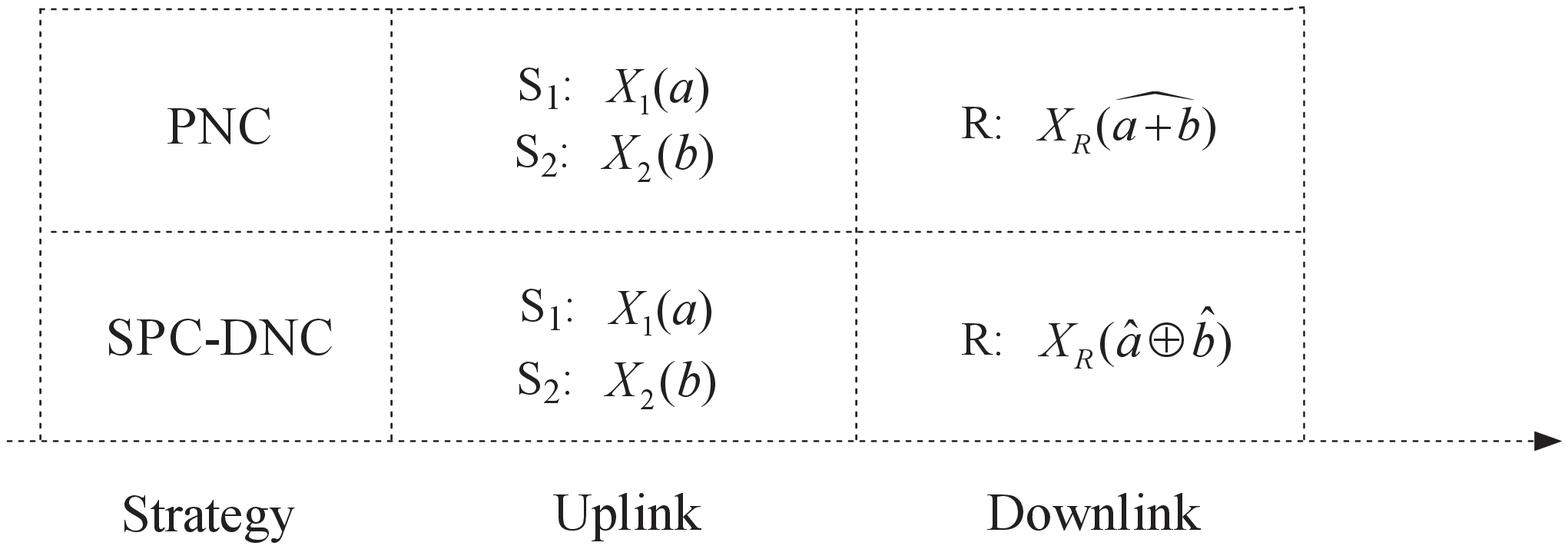}
\caption{The description of considered strategies. In PNC, a function version of $a$ and $b$ is decoded at relay and forwarded to both users. In SPC-DNC, both $a$ and $b$ are decoded at relay and then combined together before broadcasting on the downlink.
It is assumed that $a$ and $b$ are of the same length.}
\label{fig:system}
\end{figure}

In this work, we aim to minimize average energy usage for a TWRN subject to a symmetric end-to-end rate requirement from both sources, which might be required by video/audio applications.
The two considered strategies are PNC and SPC-DNC, which consist of two phases, including the multi-access uplink phase and the network-coded broadcasting downlink phase, as shown in Fig. \ref{fig:system}.

To minimize energy usage, we shall firstly review the energy usage of the two strategies, followed by the determination of the criterion rule in selection.
Finally, the PNC/SPC-DNC switching scheme is presented with the designed iterative algorithm.

\section{Power Consumption Analysis}
In this section, we shall firstly discuss the energy usage of the PNC and SPC-DNC schemes separately.
We then move on to find the rule in scheme selection for energy usage minimizing.

\subsection{PNC}
PNC is consisted of two phases. In the uplink phase, the two source nodes transmit $A$ and $B$ simultaneously to the relay node and the relay node decodes a function message $f(A+B)$ and then forward it to both sources on the downlink. As each source has complete prior knowledge of its own transmitted message, it can subtract this message and then decodes the message from the other source. In \cite{narayanan2007joint}, it is found that the achievable PNC uplink rate $R_u^{PNC}$
is given by,
\begin{align}
R_u^{PNC} = \log_2 (\frac{1}{2} + SNR)
\end{align}
where $SNR$ is the receive $SNR$ at each source node.
The required power at $S_i$ to support transmit rate $R$ on the uplink is therefore given by,
\begin{align}
P_{ui}^{PNC}=\frac{2^{R}-\frac{1}{2}}{g_{ir}} \label{eq:PNC_1i}
\end{align}
and the total required power on the uplink then is
\begin{align}
P_u^{PNC}=\sum_{i=1}^2 P_{ui}^{PNC} =\sum_{i=1}^2 \frac{2^{R}-\frac{1}{2}}{g_{ir}} \label{eq:PNC_2}
\end{align}
On the downlink, the relay node broadcasts the decoded function message to both source nodes and the minimal power required to support broadcast rate $R$ is given by,
\begin{align}
P_d^{PNC} = \frac{2^{R}-1}{g_{rm}}  \label{eq:PNC_3}
\end{align}
where $g_{rm} = \min_i g_{ri}$ follows from that the broadcast rate is determined by the minimum channel gain of all source nodes.

\subsection{SPC-DNC}
The SPC-DNC scheme time shares the traditional multi-access uplink phase and the network coded broadcasting over the downlink. On the uplink, assuming that $g_{1r} \leq g_{2r}$, from \cite{Tse05}, the messages from $S_2$ should be decoded first to minimize sum power consumption and the power of each source node is given by\footnote{Here for simplicity we assume that each source node transmits the same amount of bits to the relay node on the uplink due to the symmetric requirement. In practice, it is also beneficial in saving the buffer space at the relay node as only network-coded messages are buffered at relay for downlink transmission.},
\begin{align}
P_{u1}^{DNC}&= \frac{2^{R}-1}{g_{1r}} \\
P_{u2}^{DNC}&= \frac{\left(1+P_{{u1}}g_{1r}\right)\left(2^{R}-1\right)}{g_{2r}}
\label{eq:DNC_1}
\end{align}
and the minimal sum power required is
given by
\begin{align}
P_u^{DNC}&=\frac{2^{R}-1}{g_{1r}}+\frac{2^{R}\left(2^{R}-1\right)}{g_{2r}}
\label{eq:DNC_2}\\
&=\frac{2^{2R}-1}{g_{Mr}}+2^{R}\left(\frac{1}{g_{mr}}-\frac{1}{g_{Mr}} \right)-\frac{1}{g_{mr}}\nonumber
\end{align}
where we define $g_{Mr}=\max\{g_{1r},g_{2r}\}$ and $g_{mr}=\min\{g_{1r},g_{2r}\}$ to simplify notation.
On the downlink,
the relay node also transmits the combined messages
from the uplink and the transmit power
required is identical to that given in (\ref{eq:PNC_3}) and is omitted here.

\subsection{Switching Criterion Analysis}
Given both the power consumption for PNC and SPC-DNC, we are interested in comparing them in terms of energy usage, given the same transmit rate requirement,
The rule on selection of PNC and SPC-DNC are hence presented in Theorem \ref{the}.

\begin{thm} \label{the}
Given the channel realization and the uplink rate, PNC consumes less energy than SPC-DNC iff the following inequality holds,
\begin{align}
\frac{1}{ 2g_{mr}}
-\frac{1}{ 2g_{Mr}}
\leq \frac{2^{R}( 2^{R}-2)}{g_{Mr}}.
\label{eq:con_switch}
\end{align}
\end{thm}
\begin{IEEEproof}
It is observed that on the downlink both PNC and SPC-DNC consumes the same energy given the same broadcast rate. Hence we only need to compare the energy consumed by PNC or SPC-DNC on the uplink. Suppose the transmit rate over the uplink from both sources are $R$,
we have
\begin{align}
&\quad E_{PNC}-E_{DNC}\\&=(2^{R}-\frac{1}{2})\sum_{i=1}^2 \frac{1}{g_{ir}}-
\frac{2^{R}-1}{g_{mr}}-\frac{2^{R}(2^{R}-1)}{g_{Mr}}\\
&=\frac{2^{R}(2 - 2^{R})}{g_{Mr}}
 + \frac{1}{ 2g_{mr}}
 -\frac{1}{ 2g_{Mr}}
 \label{eq:the}
\end{align}
Hence if (\ref{eq:con_switch}) holds, we have $E_{PNC}<E_{DNC}$ and concludes that PNC is more energy-efficient than SPC-DNC and should be selected for the given channel realization and transmit rate. Otherwise, SPC-DNC uses less energy and is preferred.
\end{IEEEproof}
Further, from Theorem \ref{the} we can readily arrive at Lemma \ref{lem} as follows,
\begin{lemma}\label{lem}
If all the channel gains are equal, PNC is more energy-efficient than SPC-DNC if the following inequality holds true and less energy-efficient otherwise.
\begin{align}
R \ge 1 \label{eq:lem2}
\end{align}
\end{lemma}
The proof is omitted here as (\ref{eq:lem2}) can be readily obtained by solving a quadratic equation from (\ref{eq:the}).

Based on the observations above, it is therefore concluded that, PNC is beneficial under relatively high data requirements in terms of energy usage. On the other hand, SPC-DNC is preferred for energy usage reduction in low-SNR regime.

\section{Optimal Switching Strategy}
It is noted that both SPC-DNC and PNC consumes the same amount of energy on the downlink with the same broadcast rate. However, as observed, on the uplink,
both strategies may consume different amounts of energy under channel variations.
Therefore, it is promising to further reduce energy usage by smartly switching
among the multi-access uplink transmission of SPC-DNC and the uplink of PNC
under different channel realizations to minimize total power consumption, given the QoS requirement is met.
In this sense, we define
${P}_{u}^{\tt opt}=\min\{P_u^{PNC},P_u^{DNC}\}$, i.e., PNC or SPC-DNC are selected based on their power usage to reduce energy usage.
In addition, we denote $f_u$ and $f_d$ as the time fraction assigned to the uplink and downlink transmission, respectively,
the associated optimization problem, termed as {\bf P1}, is therefore formulated as follows,
\begin{align}
\min_{f_u,f_d,R_u,R_d} \quad f_u\bar{P}_{u}^{{\tt opt}} +f_d\bar{P}_{d} \label{eq:opt_obj}
\end{align}
subject to the following constraints,
\begin{align}
&f_u\bar{R}_{u}^{{\tt opt}} \ge \lambda, \label{eq:opt_con_up} \\
&f_d\bar{R}_{d} \ge \lambda, \label{eq:opt_con_down} \\
&f_u + f_d \le 1 \label{eq:opt_con_timesplit}
\end{align}
where $\bar{P}_{u}^{{\tt opt}}$ and $\bar{P}_{d}$ are the averaged minimal sum power usage on the uplink and that on the downlink over the distribution of the associated channel gain distributions.
(\ref{eq:opt_con_up}) and (\ref{eq:opt_con_down}) are the rate requirements for the uplink and downlink. (\ref{eq:opt_con_timesplit}) is the physical time resource splitting constraint.

Note that {\bf P1} is not a convex optimization problem, due to the quadratic terms as well as the term ${P}_{u}^{\tt opt}$, which is not convex as the minimal of two convex functions may not be convex functions.

\begin{figure}[t]
\centering
\includegraphics[width = 12cm]{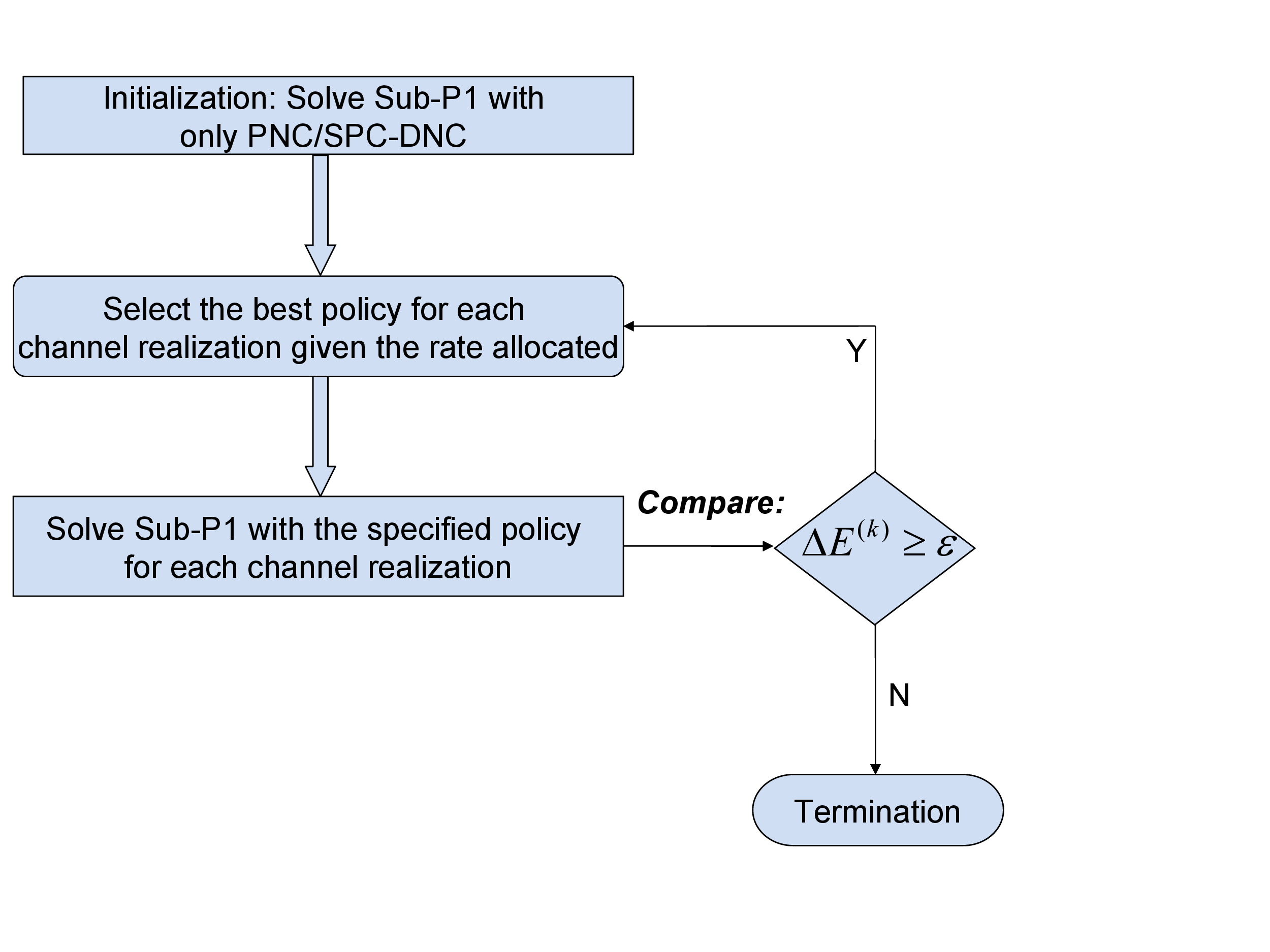}
\caption{The flowchart of the designed iterative algorithm.}
\label{fig:flowchart}
\end{figure}

Instead, we can solve {\bf P1} by firstly assuming that only PNC/DNC is used and then iteratively updating the transmit scheme in terms of energy usage for each channel gain realization. In each iterative step,
given the transmit rate allocated to each channel realization from last step, the more energy efficient strategy (PNC or SPC-DNC) is adopted and hence the energy usage is reduced at each step. The iteration ends until no additional energy reduction is attainable.
In this sense, the switching scheme must uses less energy than employing only PNC or SPC-DNC.
To be specific, in each step, the transmission strategy for each realized channel gain is determined and the associated optimization is referred to as {\bf sub-P1}.
Giving the analysis above, the steps of the algorithm is hence presented as follows.
\begin{enumerate}
\item Input: All possible $g_{ir}$, $g_{ri}$ and $\epsilon$ (a predefined threshold)
\item Initialization: solve {\bf sub-P1} by assuming that only SPC-DNC/PNC is employed in transmission and obtain the total energy usage $E^{0}$ in the initialized step.
\item Iteration: Compare the energy used in PNC and SPC-DNC given the rate allocated in the last iteration under all possible channel realizations, and use the strategy with the minimal energy usage to replace the strategy employed in the last iteration and rerun {\bf sub-P1} and obtain the associated total energy usage $E^{(k)}$.
\item Go to Step 3) if $\Delta E^{(k)}=|E^{(k)}-E^{(k-1)}|/E^{(k)} \ge \epsilon$ and go to step 5) otherwise.
\item Output: $E^{(k)}$, the optimal time splitting, the best strategy for each channel realization, the associated allocated rate as well as the transmit power level in solving {\bf sub-P1} in the $k$th iteration.
\end{enumerate}
For clarity, a flowchart of the iterative algorithm is also plotted, as shown in Fig. \ref{fig:flowchart}

\begin{remark}
Note that in each iteration step, the energy usage is reduced. In addition, the total energy usage is lower bounded by zero in nature. Combining these two facts into account, the convergence of the iterative algorithm is guaranteed and at least some local-optimal point is achieved. Further, it is worth noting that, in our implementation, the algorithm converges within tens of iterations and is hence promising in practice.
\end{remark}

\subsection{Analysis of {\bf sub-P1}}
It is also noted that at each iteration of the iterative algorithm,
{\bf P1} reduces to {\bf sub-P1} where SPC-DNC/PNC is selected and specified for each channel realization. It will be shown that {\bf sub-P1} is an equivalent convex optimization problem, followed by the analysis based on the Karush-Kuhn-Tucker (KKT) conditions, which solves {\bf sub-P1} efficiently.

To be specific, ${P}_{u}^{{\tt opt}}$ in {\bf P1} is replaced by ${P}_{u}$ in {\bf sub-P1}. In addition, to circumvent the difficulty of the quadratic terms, we define $T_a=f_a{R}_{a}$ ($a=u,d$) and $\Theta_a=f_aP_a$ ($a=u,d$). Hence, {\bf sub-P1} can be formulated as follows,
\begin{align}
\min_{f_u,f_d,T_u,T_d} \quad \bar{\Theta}_{u} +\bar{\Theta}_{d} \label{eq:opt_obj}
\end{align}
subject to the following constraints,
\begin{align}
&\bar{T}_{u} \ge \lambda, \label{eq:opt_con_up} \\
&\bar{T}_{d} \ge \lambda, \label{eq:opt_con_down} \\
&f_u + f_d \le 1 \label{eq:opt_con_timesplit}
\end{align}
where
\begin{align}
\Theta_u = \left\{
\begin{array}{ll}
\sum_{i=1}^2f_u\frac{2^{\frac{T_u}{f_u}}-1}{g_{ir}} &\mbox{for PNC};\\
f_u\frac{2^{\frac{2T_u}{f_u}}}{g_{Mr}}+
f_u 2^{\frac{T_u}{f_u}} \left( \frac{1}{g_{mr}}-\frac{1}{g_{Mr}} \right)
- \frac{f_u}{g_{mr}} & \mbox{for DNC}.
\end{array}\right.
\end{align}
and
\begin{align}
\Theta_d = f_d\frac{2^{\frac{T_d}{f_d}}-1}{g_{rm}}
\end{align}
are the the linear transformations of the perspectives of the corresponding convex functions $\left(2^R-1\right)/g$ and hence preserve convexity.

The lagrangian function associated with {\bf sub P1} is hence given by,
\begin{align}
L(T_u,T_d,f_u,f_d,\beta_u,\beta_d,\gamma)
=\bar{\Theta}_{u} +\bar{\Theta}_{d}-\beta_1\left(\bar{\Theta}_{u}-\lambda \right)-\beta_2\left(\bar{\Theta}_{d}-\lambda \right)-\gamma\left(f_u+f_d-1\right)
\end{align}
The associated KKT conditions are then derived as follows,
\begin{align}
&-\beta_1+\frac{2\ln2}{g_{Mr}}2^{\frac{2T_u}{f_u}}+\left( \frac{1}{g_{mr}}-\frac{1}{g_{Mr}} \right)2^{\frac{T_u}{f_u}}\ln2=0, \quad \mbox{DNC uplink for $P_u$.} \label{eq:downlink_opt_dnc}\\
&-\beta_1+\sum_{i=1}^{2}\frac{\ln2}{g_{ir}}2^{\frac{T_u}{f_u}}=0, \quad \mbox{PNC uplink for $P_u$.} \label{eq:downlink_opt_pnc}\\
&-\beta_2+2^{\frac{T_d}{f_d}}\frac{\ln2}{g_{rm}}=0, \quad \mbox{downlink for $P_d$.} \label{eq:downlink_opt} \\
&\frac{2^{\frac{2T_u}{f_u}}}{g_{Mr}}+
 2^{\frac{T_u}{f_u}} \left( \frac{1}{g_{mr}}-\frac{1}{g_{Mr}} \right)
- \frac{1}{g_{mr}}
-\frac{2T_u\ln2}{f_u}\frac{2^{\frac{2T_u}{f_u}}}{g_{Mr}} \nonumber\\
&-\frac{T_u\ln2}{f_u}2^{\frac{T_u}{f_u}} \left( \frac{1}{g_{mr}}-\frac{1}{g_{Mr}} \right)
+\gamma=0, \quad  \quad \mbox{DNC uplink for $f_u$} \label{eq:downlink_opt_dnc_f}\\
&\sum_{i=1}^2\frac{2^{\frac{T_u}{f_u}}-1}{g_{ir}}-\sum_{i=1}^2\frac{2^{\frac{T_u}{f_u}}}{g_{ir}}\frac{T_u \ln 2}{f_u}+\gamma=0 \quad \mbox{PNC uplink for $f_u$} \label{eq:downlink_opt_pnc_f}\\
&\frac{2^{\frac{T_d}{f_d}}-1}{g_{rm}}-\frac{T_d \ln 2}{f_d}\frac{2^{\frac{T_d}{f_d}}}{g_{rm}}+\gamma=0 \quad \mbox{downlink for $f_d$} \label{eq:downlink_opt_f}
\end{align}

After some arithmetic operations, the optimal power allocations for the uplink is then derived as follows,
\begin{align}
P_u = \left\{
\begin{array}{ll}
\sum_{i=1}^{2}\left( \beta_1 \log_2e \frac{g_{3-i,r}}{\sum_{i=1}^2g_{ir}} -\frac{1}{g_{ir}}   \right)^+,  &\mbox{for PNC};\\
 \left( \frac{x-1}{g_{mr}}\right)^+ + \left( \frac{x\left(x-1\right)}{g_{Mr}}\right)^+, & \mbox{for DNC}.
\end{array}\right.
\end{align}
where
\begin{align}
x=\frac{1}{4}\left(1-\frac{g_{Mr}}{g_{mr}}\right)
+\frac{1}{4\ln2}
\sqrt{\left(1-\frac{g_{Mr}}{g_{mr}}\right)^2
+4\beta_1\left(g_{Mr}-\frac{g_{Mr}^2}{g_{mr}}\right)}
\end{align}

The downlink optimal power allocations with respect to the channel gains can be similarly derived from (\ref{eq:downlink_opt}) and is presented below.
\begin{align}
P_d=\left( \beta_2 \log_2e -\frac{1}{g_{rm}}   \right)^+
\end{align}

For the optimal time splitting, it is noted that the closed-form solutions are not tractable as the associated KKT conditions are transcendental equations. However, numerical algorithms can be applied to find the optimal $f_u$ and $f_d$. Henceforth, {\bf sub-P1} can be solved efficiently by KKT conditions and in each iteration of the presented algorithm the global optimal solution is obtained and hence we argue that the proposed algorithm in Sec. V above leads to a sub-optimal solution.

\begin{figure}[t]
\centering
\includegraphics[width = 12cm]{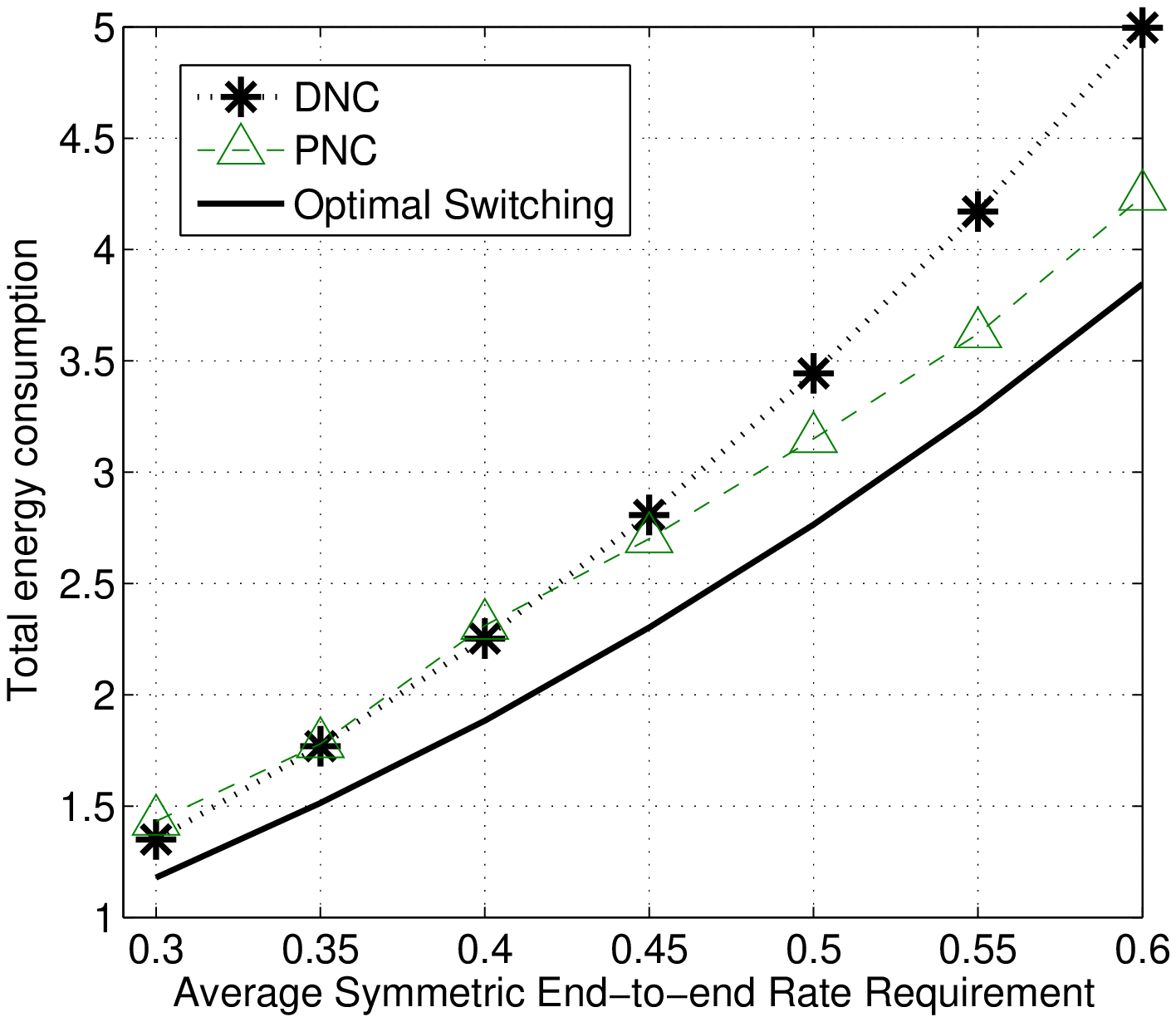}
\caption{The total power usage required of different schemes subject to the symmetric end-to-end rate requirement.}
\label{fig:all}
\end{figure}

\section{Numerical Results}
In this section, numerical results are presented to verify our findings. In the considered setting, noise at each node is assumed to be Gaussian with zero mean and unit variance and all links are assumed to be Rayleigh fading channels with unity link gain on average. The reciprocity of the associated uplink and downlink channels is assumed.
The average total energy usage of each scheme is obtained by averaging over $1000$ independent realizations of link gains and the average symmetric end-to-end rate requirement on both sides is in the unit of bit/s/Hz.
As observed in Fig. \ref{fig:all},  PNC performs better than SPC-DNC with relatively high data rate requirement and worse with low data rate requirement. In addition, it is observed that the optimal switching scheme outperforms solely PNC and SPC-DNC schemes for all data rate requirements and the performance gain achieved is hence demonstrated, validating the superiority of our designed switching scheme.

\section{Conclusion}
In this work, we studied a three-node, two-way relaying system. Our aim was to minimize average total energy usage for a TWRN by switching between PNC and SPC-DNC, while satisfying the QoS requirement. To this end, we analytically derived the optimal selection criterion for SPC-DNC and PNC for each channel gain realization and the associated optimal problem to minimize energy usage by switching between SPC-DNC and PNC was formulated and solved. The performance gain of the designed adaptable PNC/DNC switching scheme, over the schemes by only employing SPC-DNC or PNC, was validated by numerical results.



\end{document}